\documentclass[%
superscriptaddress,
 aps,
 reprint
]{revtex4-2}

\usepackage{graphicx}% Include figure files
\usepackage{dcolumn}% Align table columns on decimal point
\usepackage{bm}% bold math
\usepackage{amsmath}
\usepackage{tablefootnote}
\usepackage{array,booktabs,siunitx}
\usepackage{tabularx}
\usepackage{xcolor}
\usepackage[utf8]{inputenc}
\usepackage[T1]{fontenc}
\usepackage{mathptmx}
\usepackage{float}
\usepackage{sidecap}
\usepackage{siunitx}
\usepackage{gensymb}

\begin{document}

\title[]{Johnson-noise-limited cancellation-free microwave impedance microscopy \\with monolithic silicon cantilever probes}

\author{Jun-Yi Shan}
\author{Nathaniel Morrison}
\affiliation{Department of Physics, University of California, Berkeley, Berkeley, CA 94720, USA}
\affiliation{Lawrence Berkeley National Laboratory, Berkeley, CA 94720, USA}

\author{Su-Di Chen}
\affiliation{Department of Physics, University of California, Berkeley, Berkeley, CA 94720, USA}
\affiliation{Lawrence Berkeley National Laboratory, Berkeley, CA 94720, USA}
\affiliation{Kavli Energy NanoScience Institute, University of California, Berkeley, Berkeley, CA 94720 USA}

\author{Feng Wang}
\affiliation{Department of Physics, University of California, Berkeley, Berkeley, CA 94720, USA}
\affiliation{Lawrence Berkeley National Laboratory, Berkeley, CA 94720, USA}
\affiliation{Kavli Energy NanoScience Institute, University of California, Berkeley, Berkeley, CA 94720 USA}

\author{Eric Y. Ma}
\email{eric.y.ma@berkeley.edu}
\affiliation{Department of Physics, University of California, Berkeley, Berkeley, CA 94720, USA}
\affiliation{Lawrence Berkeley National Laboratory, Berkeley, CA 94720, USA}
\affiliation{Department of Electrical Engineering and Computer Sciences, University of California, Berkeley, Berkeley, CA 94720, USA\looseness=-1}

\begin{abstract}

Microwave impedance microscopy (MIM) is an emerging scanning probe technique for nanoscale complex permittivity mapping and has made significant impacts in diverse fields from semiconductors to quantum materials. To date, the most significant hurdles that limit its widespread use are the requirements of specialized microwave probes and high-precision cancellation circuits. Here we show that forgoing both elements not only is feasible but actually enhances MIM performance. Using monolithic silicon cantilever probes and a cancellation-free architecture, we demonstrate thermal Johnson-noise-limited, drift-free MIM operation with 15 nm spatial resolution, minimal topography crosstalk, and an unprecedented sensitivity of 0.26 zF/$\sqrt{\text{Hz}}$. We accomplish this by taking advantage of the high mechanical resonant frequency and spatial resolution of silicon probes, the inherent common-mode phase noise rejection of self-referenced homodyne detection, and the exceptional stability of the streamlined architecture. Our approach makes MIM drastically more accessible and paves the way for more advanced operation modes and integration with complementary techniques. 
 
\end{abstract}

\maketitle

\section{Introduction}

Microwave impedance microscopy (MIM) allows nanoscale mapping of local complex permittivity in a wide variety of systems under diverse conditions \cite{lai,barber,berweger2020nanoelectronic,hoffmann2014measuring,huber2012calibrated}, from buried 2D electronic systems at mK temperatures and topological materials in strong magnetic fields, to semiconductor devices under optical illumination and biological samples in liquid \cite{jiang2023implementing,cao2023millikelvin,chen2022excitonic,allen2019visualization,chu2020unveiling,gramse2017nondestructive,ohlberg2021limits,tselev2016seeing}. MIM achieves this by delivering microwave fields to a sharp metallic tip and measuring the minuscule changes in the complex reflection coefficient due to near-field tip-sample interaction.

To date, two key elements are believed to be essential to MIM: specialized probes for microwave delivery that minimize stray coupling, probe capacitance, and ohmic loss; and a high-precision cancellation circuit that removes the portion of the reflected microwave unperturbed by tip-sample interaction via destructive interference, thereby allowing high gain to achieve the required sensitivity (Fig. 1a). 

Nevertheless, these two elements introduce significant complexity and cost, raising barriers to broader adoption and, more importantly, inadvertently compromising the best achievable performance of MIM. The specialized microwave probes -- ranging from stripline cantilevers requiring intricate micro-electromechanical fabrication \cite{yang2012} to manually etched and glued long-shank bare metal wires and cantilevers \cite{karbassi2008quantitative,huber2010calibrated} -- typically provide only moderate tip sizes with low reliability and reproducibility compared with monolithic silicon probes. Furthermore, they exhibit relatively low mechanical resonant frequencies and quality factors ($Q$-factors). Such limitations constrain the ultimate topography performance, introduce artifacts, and hinder the consistent, quantitative interpretation of MIM signals.

Furthermore, the cancellation circuits, typically comprising multiple stages of mechanical and voltage- or digitally-controlled amplitude attenuators and phase shifters, are prone to external disturbance. Vibrations and electromagnetic interference can inject noise directly into the un-amplified MIM signal. Ambient temperature drifts and device aging lead to a fluctuating baseline that ultimately saturates the amplifier chain, limiting integration times and thereby restricting achievable sensitivity. Additionally, the cancellation circuits often contain bandwidth-limiting components. 

Here we present a Johnson-noise-limited, drift-free MIM with 15 nm electrical spatial resolution, minimal topography crosstalk, and a sensitivity of 0.26 zF/$\sqrt{\text{Hz}}$ using monolithic silicon probes and a cancellation-free architecture. We describe how we bypass the need for specialized probes or cancellation, and in turn, advance the state of the art in MIM performance. 
% combining the high mechanical resonant frequency and $Q$-factor of silicon probes with the inherent common-mode phase noise rejection of self-referenced homodyne detection, together with optimized system design and integration, which in turn allowed us to advance the state of the art. 
We discuss how our new approach facilitates advanced MIM operation such as spectroscopy and nonlinear mode, and seamless integration with complementary techniques that utilize high-performance silicon-based probes such as near-field optical, magnetic force, and acoustic force microscopy \cite{zhang2013quantitative,martin1987magnetic,rabe1994acoustic}.

\begin{figure*}
\includegraphics[width=1\textwidth]{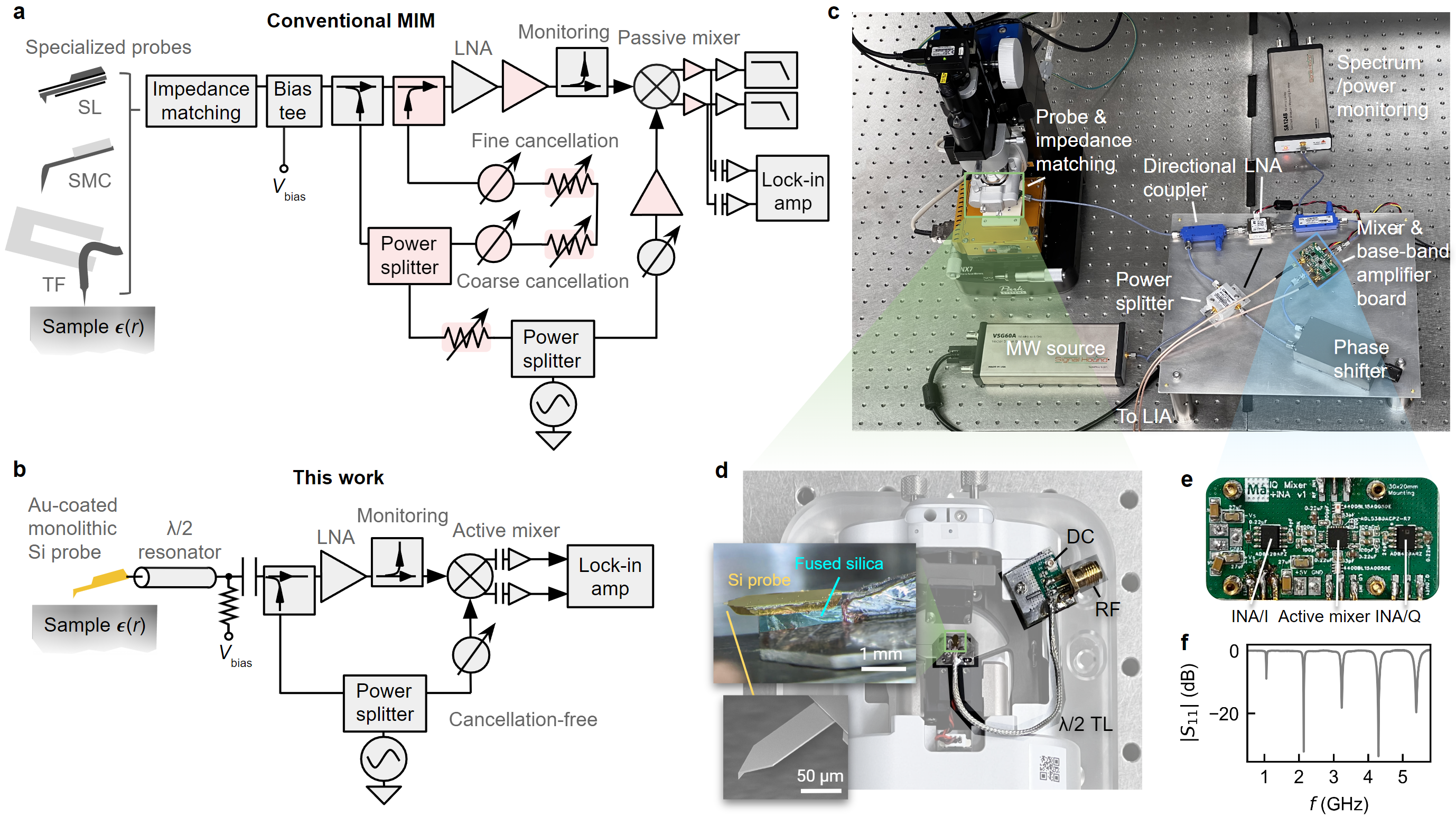}% Here is how to import EPS art
\caption{\label{fig:1} \textbf{Cancellation-free MIM with monolithic silicon cantilever probes.} \textbf{a, b}, Design of conventional MIM (\textbf{a}) and this work (\textbf{b}). Components removed in our streamlined architecture are shaded red. \textbf{c}, Implementation of our design. \textbf{d}, Details of the probe and impedance matching network attached to the atomic force microscope (AFM) head. Inset: Probe mounting details and scanning electron micrograph of the Si probe, from \cite{160acgg}. \textbf{e}, The custom printed circuit board integrating the active mixer and INAs. \textbf{f}, $|S_{11}|$ spectrum of the Si probe impedance matched by a half-wave resonator.}
\end{figure*}

\section{Results}
\subsection{Design and Validation} 
We first briefly review the principle and design considerations of conventional MIM (Fig. 1a). A source generates microwave which is split into three parts. The first part is delivered to specialized MIM probes through a directional coupler and an impedance matching network \cite{shan2022universal,shan2023circuit}. The second part passes through multiple stages of tunable attenuators and phase shifters, and is combined with the microwave reflected from the tip-sample in order to cancel the baseline reflection through destructive interference. The baseline-free signal then gets amplified through several low-noise amplifiers (LNA), and mixed with the third microwave channel in order to down-convert the signal to baseband frequencies. An additional directional coupler is often inserted between the LNA and the mixer for monitoring. The resulting in-phase (I) and quadrature (Q) signals are then amplified by baseband amplifiers and detected after a low-pass filter, or, in dynamic mode, through a lock-in amplifier (LIA) locked to the probe's mechanical vibration or other modulations.

Our design differs from the above in several important aspects, with the most significant being the adoption of standard gold-coated monolithic silicon cantilever probes, and the complete elimination of the cancellation circuits (Fig. 1b, c). In addition, we opted for a single LNA and an active mixer, and directly AC-coupled the I and Q outputs before further amplification through high-gain instrumentation amplifiers (INA). We integrated the active mixer and INAs onto a single custom printed circuit board (PCB) to minimize signal loss and noise (Fig. 1e). See Methods section "MIM electronics and probes" for complete details.

Defying the conventional wisdom, our MIM implementation is not only operational but also high performing. For an initial evaluation, we used an Al dot sample, featuring 5 $\mu$m $\times$ 5 $\mu$m $\times$ 15 nm Al squares patterned on SiO$_2$. Figure 2 presents a side-by-side comparison of images taken with our approach and with a state-of-the-art commercial MIM system using multi-stage cancellation and a solid metal cantilever (SMC) probe, at identical microwave power and scan rate. Both sets correctly show strong contrasts between Al and SiO$_2$ in the reactive channel (MIM-Im) and minimal contrast in the dissipative channel (MIM-Re) \cite{barber,chu2020microwave}. The lower MIM-Im signal associated with surface particulates on Al proves that the MIM signals originated from local conductivity instead of topography crosstalk. The same dot was imaged, and the sharper Si tip gave rise to a more faithful representation of the surface features. Despite a higher spatial resolution, which generally leads to trade-offs in MIM signal, our implementation shows an excellent signal-to-noise ratio (SNR), as evidenced by the scan line profiles (Fig. 2h).

Below we first examine the factors behind this surprising result, addressing the viability of Si probes and the implications of eliminating cancellation, and then show how our approach elevates the state of the art in noise, sensitivity, and spatial resolution.

\begin{figure}
\includegraphics[width=0.5\textwidth]{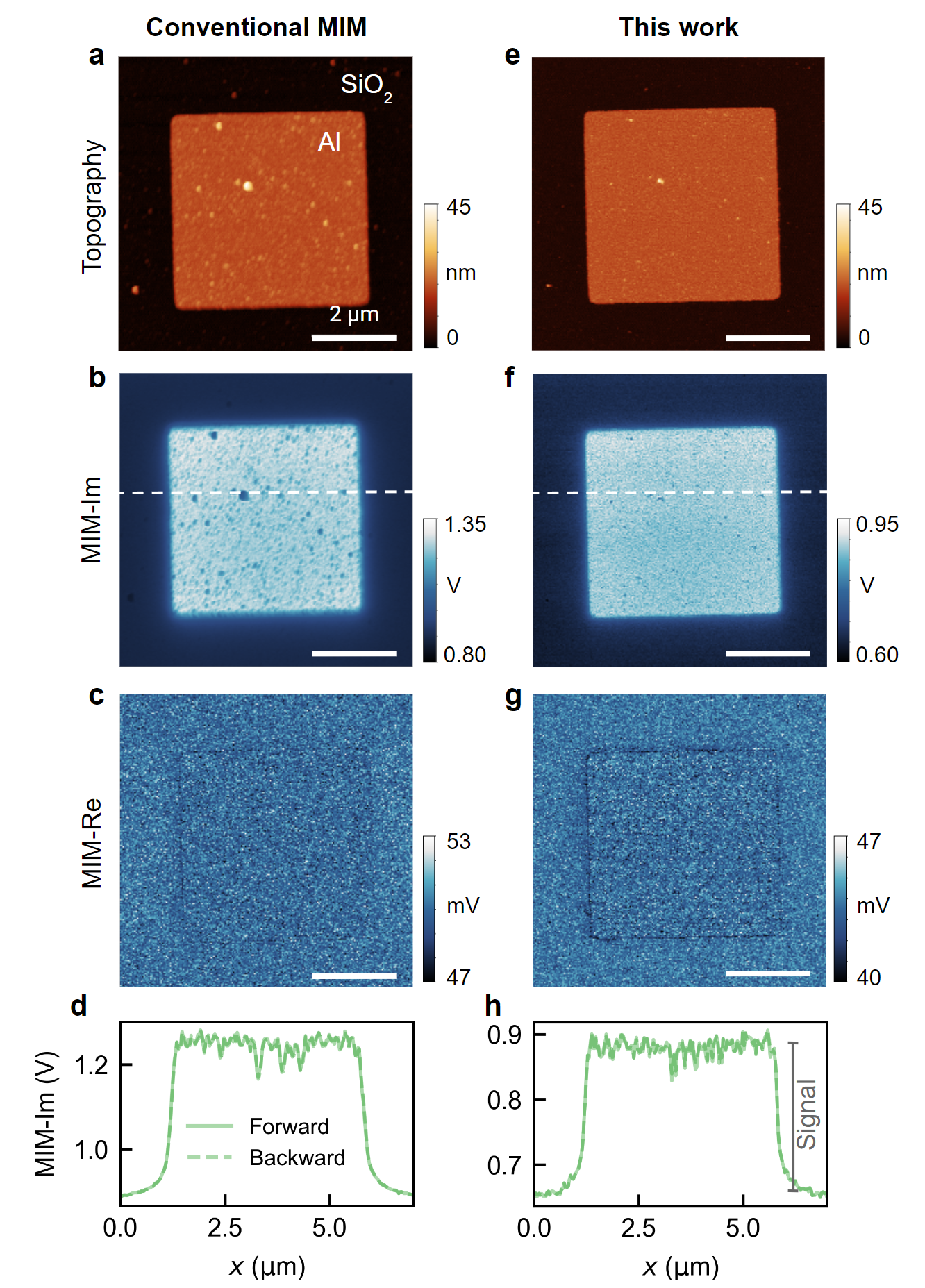}% Here is how to import EPS art
\caption{\label{fig:2} \textbf{Validation and comparison with state-of-the-art conventional system.} Al dot on SiO$_2$ measured by conventional MIM: \textbf{a}, surface topography, \textbf{b}, raw MIM-Im image, \textbf{c}, raw MIM-Re image, and \textbf{d}, line profiles along the horizontal line in panel (\textbf{b}) showing forward and backward scans. \textbf{e}-\textbf{h}, The same set of measurements taken with our MIM implementation. For both sets, the microwave frequency is 3.2 GHz, $P_\mathrm{in}=-10$ dBm, and lock-in time constant is 4 ms.}
\end{figure}

\subsection{Monolithic silicon probes for microwave}
Metal-coated monolithic silicon cantilever probes, despite their excellent force sensitivity, spatial resolution, robustness, and versatility, have generally been considered unsuitable for microwave scanning probe techniques due to perceived challenges associated with strong stray coupling, large self-capacitance, and high ohmic loss \cite{yang2012,yang2014}. 

We show that these perceived limitations are largely overestimated or can be substantially mitigated, and are thus outweighed by the benefits. First, the utilization of a dynamic mode, wherein the MIM signal is modulated by the nm-amplitude mechanical vibration of the cantilever and discerned by a lock-in amplifier, effectively nullifies stray coupling between the probe body and the sample \cite{lai2009tapping}, as corroborated by the high-resolution drift-free MIM images with minimal topographic interference shown in Fig. 2 and 4. 

Second, our modeling and characterizations show that the GHz-range dissipation and capacitance of gold-coated Si probes are moderate relative to their Pt SMC or Al stripline (SL) counterparts ($\sim2$~$\Omega$/0.1~pF vs. 0.2~$\Omega$/0.03~pF or 5~$\Omega$/1~pF, see Methods). Moreover, strategic placement of a ground plane transforms the probe body into a short microstrip waveguide, where self-capacitance is compensated by self-inductance to render a characteristic impedance close to 50 $\Omega$ (Fig. 1d inset, see also Methods). Consequently, the Si probes can be reliably impedance matched by, e.g., a simple half-wave transmission line resonator (Fig. 1d, f) to obtain a responsivity ($\eta=V_\text{tip}/V_\text{in}$) \cite{shan2022universal, shan2023circuit} on par with SMC probes (Fig. 2). This allows us to unlock key benefits of Si probes, such as high mechanical resonant frequency, robust topography feedback, and compatibility with other complementary techniques.

\begin{figure}
\includegraphics[width=0.5\textwidth]{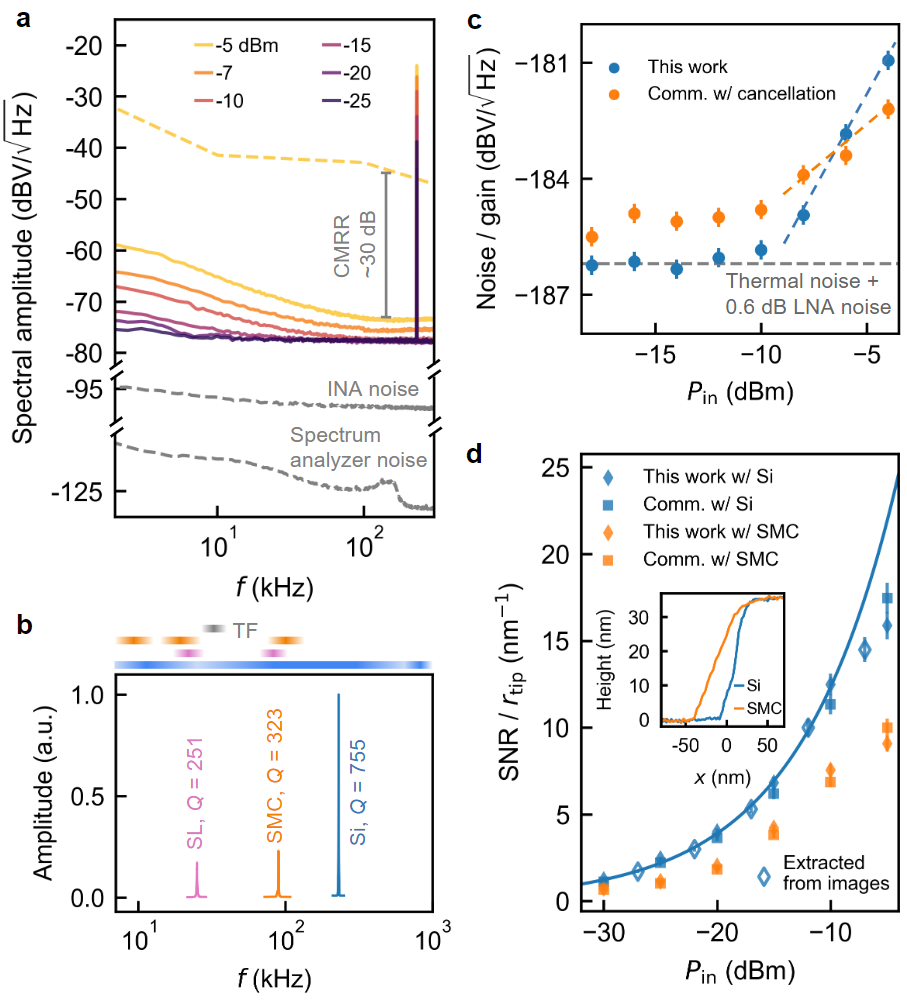}% Here is how to import EPS art
\caption{\label{fig:3} \textbf{Demonstration of Johnson-noise-limited performance.} \textbf{a}, Baseband spectral amplitude of MIM-Im signal for various $P_\mathrm{in}$ with 3.2 GHz microwave for a probe $|S_{11}|$ of -18 dB. The yellow dashed curve shows the estimated phase noise density without common-mode rejection for $P_\mathrm{in}=-5$ dBm. The grey dashed curves show the INA and the spectrum analyzer noise floor. CMRR: common-mode rejection ratio. \textbf{b}, Mechanical vibration amplitude as a function of driving frequency for the stripline (SL), solid metal cantilever (SMC), and Si probes, at the same driving power. The horizontal bars on top indicate the range of available mechanical resonant frequencies $f_0$ for tuning fork (grey), SMC (orange), SL (pink), and Si probes (blue). \textbf{c}, The noise spetral amplitude near $f_0$ normalized by total system gain as a function of $P_\mathrm{in}$ for our implementation and a commercial system with cancellation. Colored dashed lines are guides to the eye. Gray dashed line shows 295 K Johnson noise plus 0.6 dB LNA noise. \textbf{d}, Tip-size-normalized SNR as a function of $P_\mathrm{in}$ for four combinations of MIM electronics and probes. Solid and open symbols are derived from MIM signal spectra and images, respectively. The solid curve is a fit of the filled blue diamonds at $P_\mathrm{in}<-10$ dBm to $A10^{P_\mathrm{in} (\mathrm{dBm})/20}$, where $A$ is the fitting parameter. Inset: topography line cuts using two types of probes across a sharp edge to determine the tip radius.}
\end{figure}

\subsection{Cancellation-free operation}
We now discuss why cancellation-free operation is possible. The first challenge is that the baseline reflection, as large as $\sim-20$ dBm, could easily saturate the signal chain that requires $\sim100$ dB of total gain to detect the minuscule signal. We overcome this by reducing microwave gain, operating in dynamic mode, and removing the baseline \textit{after} down-conversion. More specifically, we use a single LNA before the mixer and directly AC-couple the down-converted mixer outputs. This way, the DC component (< 1 kHz) that represents the baseline is rejected, while AC components at the cantilever vibration frequency $f_0$, representing the dynamic-mode MIM signal, are allowed to pass and amplified further by high-gain (66 dB) INAs (Fig. 1b). The signal chain never saturates as long as the reflected power is less than $\sim-20$ dBm, a condition almost always satisfied with proper impedance matching.   

Phase noise from the baseline reflection is another potential challenge, and could seemingly overwhelm the dynamic MIM signal regardless of whether the DC component is removed. For instance, our standard microwave source has a phase noise figure of -115 dBc/Hz at 100 kHz offset from 3 GHz, which would imply a voltage noise density of -44 dBV/$\sqrt{\text{Hz}}$ near $f_0=230$ kHz in the down-converted baseband signal ($P_\text{in}=-5$ dBm, $|S_{11}|=-18$ dB, total system gain = 110 dB, yellow dashed curve in Fig. 3a). Such noise would overpower all other noise sources and drastically diminish the SNR, precluding cancellation-free operation. Yet, our measured noise density is more than 30 dB lower (yellow solid curve), making high-SNR MIM feasible without cancellation. 

This unexpected outcome stems from the intrinsic common-mode phase noise rejection of \textit{self-referenced} homodyne detection \cite{phasenoise}. As the signal and the reference are derived from the same source, they contain the same phase noise, which effectively cancels during mixing. For a signal voltage $V_\mathrm{sig}(t)=V_1\cos[\omega t+\phi(t)]$ and a reference voltage $V_\mathrm{ref}(t)=V_2\cos[\omega t+\phi_0+\phi(t)]$, their mixing result is $V_\mathrm{sig}(t)\cdot V_\mathrm{ref}(t) = (1/2)V_1V_2[\cos[2\omega t+2\phi(t)+\phi_0]+\cos\phi_0]$, in which the baseband component is independent of the phase noise $\phi(t)$. Experimentally, this common-mode rejection is finite ($\sim30$ dB in our setup) due to cable length differences, amplifier phase noise, and phase-to-amplitude noise conversion \cite{milotti1998amplitude}. Nevertheless, it allows additive broadband noise -- dominated by Johnson noise, as we will demonstrate -- to become the primary contributor at low to moderate power levels (Fig. 3a).

\subsection{Johnson-noise-limited noise floor}
We now show that our MIM implementation achieves Johnson-noise-limited noise floor. In MIM measurements, electronic noise arises from two primary sources: noise that scales with the total reflected power such as the phase noise mentioned above, and additive power-independent noise, which includes Johnson-Nyquist noise, amplifier broadband noise, and electromagnetic interference. The input-referenced Johnson noise density is $k_BT$, where $k_B$ is the Boltzmann constant and $T$ is temperature. Fig. 3a illustrates that while residual phase noise dominates at input powers above -7 dBm near $f_0=230$ kHz, the noise floor converges to a nearly flat profile, not limited by the spectrum analyzer or INA noise, for $P_\text{in}\le-10$ dBm. Meanwhile, the dynamic MIM signal at $f_0$ scales linearly with $P_\text{in}$ down to -25 dBm (Extended Data Fig. 1). 

By plotting the gain-normalized noise density at $f_0$ against various $P_\text{in}$ levels, we confirm Johnson-noise-limited operation for $P_\text{in}\le-10$ dBm (Fig. 3c). Within this regime, the Johnson noise near the GHz carrier frequency, emanating from the impedance-matched probe at 295 K is -186.8 dBV/$\sqrt{\text{Hz}}$ for 50 $\Omega$ and is slightly intensified by the LNA noise (0.6 dB), dominating the measured noise floor after amplification and down-conversion. The LNA's 36 dB gain ensures that noise contributions from downstream components are negligible \cite{haus1958optimum}. To further lower the noise floor, one would need to lower the temperature of both the probe and the LNA.  

To our knowledge, this is the first demonstration of Johnson-noise-limited MIM operation. We attribute this milestone to the high mechanical resonant frequency of Si probes and the significantly streamlined architecture. As shown in Fig. 3a, the maximal power to maintain Johnson-noise-limited operation will decrease with lower probe mechanical resonance frequencies (Fig. 3b) due to elevated phase noise at reduced offset frequencies. On the other hand, while cancellation can mitigate phase noise at higher power levels, it inadvertently introduces additional noise. This noise, directly injected into the signal before the initial low-noise amplification, is likely responsible for an observed higher noise floor in the commercial system -- up to 2 dB above the Johnson noise, at low and moderate power levels (Fig. 3c).  

% The implications of this finding are twofold: first, it underscores the nuanced trade-offs involved in the design of MIM systems, and second, it suggests that the traditional emphasis on cancellation may be reconsidered in light of alternative strategies that prioritize intrinsic noise reduction and system simplification.

The Johnson-noise-limited noise floor, coupled with high responsivity, gives rise to excellent performance. Using the SNR normalized by the tip radius as a figure of merit \cite{lai2008modeling}, we compare our approach with various alternate configurations (Fig. 3d). The combination of Si probes and cancellation-free architecture consistently outperforms others in the Johnson-noise-limited regime, where signal increases linearly with $P_\mathrm{in}$ while noise remains constant (solid curve in Fig. 3d). Moreover, the excellent agreement between SNR values derived from MIM signal spectra and actual images (solid and open diamonds, see also Methods) confirms that high SNR persists in practical scanning scenarios with other potential noise variables such as fluctuating tip conditions or vibration amplitudes. This robustness lays the groundwork for achieving the unprecedented sensitivity we now discuss. 

% signal as the contrast within MIM-Im between Al and SiO$_2$ (Fig. 2h), and characterize the broadband noise either through the spectrum analyzer or by extracting from the MIM-Im image (see Methods section "Characterizing SNR"). We note that from topography (Fig. 3d inset) we estimate the tip radius $r_\mathrm{tip}$ to be 15 nm for the Si probe and 30 nm for the SMC probe, so to fairly compare the SNR across probes of difference tip sizes, we normalize the SNR by the tip radius \cite{lai2008modeling}. Figure 3d shows the normalized SNR for various combinations between the MIM electronics and the probes. Consistent with the discussions above of noises, our cancellation-free MIM implementation gives rise to higher SNR at moderate to low powers, but lower SNR when phase noise dominates the measurements. We note that in the thermally-limited regime, increasing $P_\mathrm{in}$ does not affect noise and SNR scales linearly with $P_\mathrm{in}$ (the solid curve in Fig. 2d). 

\subsection{Sub-zeptofarad (zF) sensitivity}

The excellent tip-radius-normalized SNR, when combined with the exceptional stability of silicon probes and the cancellation-free architecture, allows long pixel dwell times to achieve sub-zF sensitivity while maintaining high spatial resolution. To demonstrate this, we used an encapsulated graphene sample consisting of a monolayer graphene flake sandwiched between layers of hexagonal boron nitride (hBN) (Fig. 4a, b). This configuration allows us to again highlight our true electrical sensitivity with minimal interference from topography. The stark electrical contrast between graphene and hBN takes place across a topographically flat region, whereas the two hBN-only regions show consistent MIM signal despite a 35 nm height difference (Fig. 4c, d).  

\begin{figure}
\includegraphics[width=0.5\textwidth]{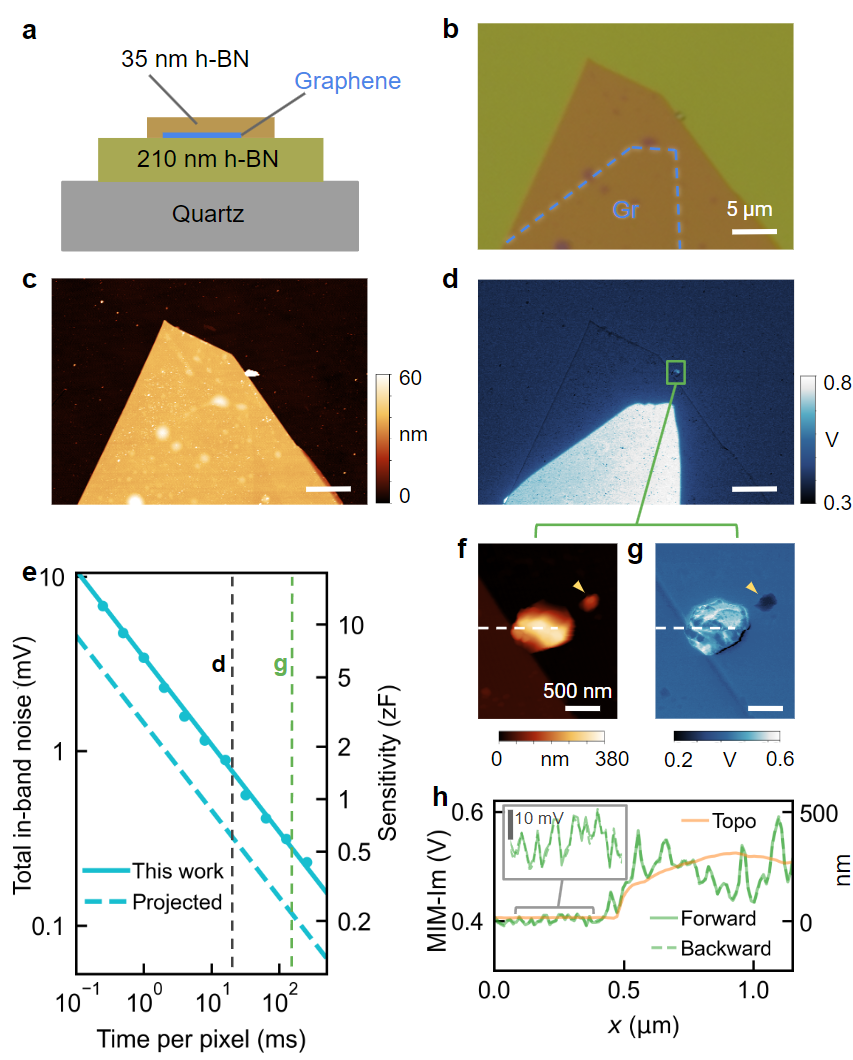}% Here is how to import EPS art
\caption{\label{fig:4} \textbf{Achieving zF sensitivity.} \textbf{a}, Structure of the encapsulated graphene sample. \textbf{b}, Optical image of the sample. The blue dashed line outlines the graphene region. \textbf{c}, Topography and \textbf{d}, raw MIM-Im image of the sample. \textbf{e}, Measured total in-band noise and the corresponding sensitivity as a function of pixel dwell time (circular markers). The sloped lines are theoretical estimates for this work (solid) and a finer proposed measurement at 77 K with lower source phase noise (dashed, see Discussion). The dashed vertical lines indicate the conditions for panels (\textbf{d}) and (\textbf{g}). The noise density at room temperature for this work is 0.14 mV/$\sqrt{\text{Hz}}$. \textbf{f}, A zoomed-in topography image of (\textbf{d}). \textbf{g}, Raw MIM-Im image of the same region taken over 6 hours. The smaller feature (yellow arrow) appears insulating with a lower permittivity than hBN and is likely a dust particle; the larger feature appears highly conductive and is likely a mixture of graphene fragments and polymer residue that was pushed out of the top hBN during fabrication. \textbf{h}, Line profiles from (\textbf{f, g}) with a zoomed-in view highlighting the SNR. }
\end{figure}

We then demonstrate an unprecedented sensitivity $\sim0.5$ zF. In principle, sensitivity -- defined as the signal corresponding to an SNR of 1 -- can be enhanced simply by narrowing the measurement bandwidth, thereby reducing the total in-band noise (Fig. 4e). In our dynamic-mode MIM setup, this involves increasing the lock-in time constant and the pixel dwell time to ensure signal stabilization. For instance, we used a lock-in time constant of 4 ms and a pixel dwell time of 16 ms for Fig. 4d, translating to a total in-band noise of 0.17 mV and a sensitivity of 1.7 zF, based on a measured responsivity (see Methods). Practically, however, one is often constrained by cancellation-induced drifts and saturation, as well as challenges like tip degradation and contaminant accumulation, particularly under ambient conditions. Our implementation substantially alleviates these issues, enabling us to further improve the sensitivity.

As an illustrative case, we imaged a 2 $\mu$m $\times$ 2.5 $\mu$m area containing two surface particulates (Fig. 4f, g). Using a lock-in time constant of 40 ms and a pixel dwell time of 160 ms, we achieved a sensitivity of 0.53 zF, or a bandwidth normalized sensitivity of 0.26 zF/$\sqrt{\text{Hz}}$, representing a 3 - 4x improvement over previously reported figures in the related scanning capacitance microscopy (SCM) technique \cite{tran2001zeptofarad,dongmo2007sub}. We note that MIM significantly differs from SCM due to its simultaneous quadrature sensitivity and minimal sample preparation requirements. The 6-hour scan was conducted in an open environment without enclosure or active temperature stabilization -- a task inconceivable with conventional MIM. Under identical settings, the commercial system drifted into saturation in less than 2 hours. Remarkably, the resulting \textit{raw} MIM-Im image is virtually drift- and noise-free, has high spatial resolution, reveals detailed permittivity features uncorrelated with topography, and demonstrates excellent consistency between the forward and backward scans (Fig. 4h). This accomplishment shows the possibility of approaching even higher sensitivity, a prospect we explore in subsequent discussions. 

\begin{figure*}
\includegraphics[width=1\textwidth]{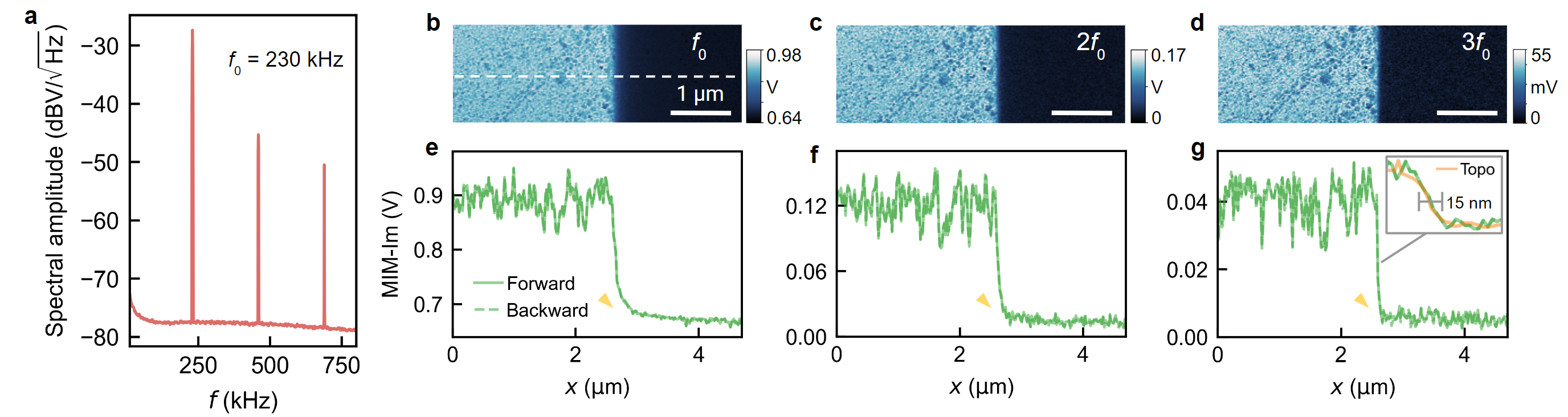}% Here is how to import EPS art
\caption{\label{fig:5} \textbf{Enhanced spatial resolution through harmonic demodulation.} \textbf{a}, The measured MIM signal spectral amplitude with $P_\mathrm{in}=-10$ dBm over a wider frequency range, showing dynamic MIM signal at harmonics of $f_0$. \textbf{b}-\textbf{d}, Raw MIM-Im image demodulated at $f_0$, $2f_0$, and $3f_0$, respectively. \textbf{e}-\textbf{g}, Line profiles along the dashed line in the images with a zoomed-in view across the Al/SiO$_2$ edge for 3$f_0$ accompanied by the corresponding topography. The yellow arrows indicate the disappearance of the "halo" artifact.}
\end{figure*}

\subsection{Enhanced spatial resolution through harmonic demodulation}

Capitalizing on the robust high-resolution capabilities of silicon probes, we have achieved additional enhancements in electrical spatial resolution by demodulating the MIM signal at higher harmonics of the cantilever's mechanical resonance (Fig. 5). Contrary to short-ranged interactions like atomic force or tunneling current \cite{giessibl2005afm}, the near-field microwave interaction is relatively long-ranged due to the power-law decay of electric fields \cite{gramse2014calibrated}. As a consequence, despite the spatial contrast predominantly originating from the tip apex, contributions from the upper sections of the probe can introduce lower-resolution elements, particularly between large regions with high permittivity contrast. This phenomenon is often manifested as subtle "halos" at the boundaries between metallic and insulating regions (Figs. 2, 4, and 5b, e).

We show that demodulating at harmonics of the cantilever mechanical resonant frequency $f_0$ can effectively remove these lower-resolution features, enhancing the electrical spatial resolution and simplifying image interpretation. As illustrated in Fig. 5a, dynamic MIM signals are detectable at multiples of $f_0$, albeit with diminishing amplitude. We observe that demodulating at higher harmonics provides crisper transitions at the aluminum/SiO$_2$ boundary, eliminating the "halo" and achieving tip-radius-limited 15 nm \textit{electrical} spatial resolution at 3$f_0$ while maintaining decent SNR (Fig. 5g inset). The relative contrast from surface contaminants is also enhanced, as the signal becomes increasingly localized.

This technique works because it captures the nonlinear dependence of MIM signal to the tip-sample distance -- a response primarily dictated by the interactions near the tip apex. However, this refinement in spatial resolution comes at the expense of SNR. This trade-off can be mitigated by extending the scan duration, as previously discussed, or increasing the input power, since the harmonics are at higher offset frequencies (Fig. 3a). Notably, while similar methods have been widely used in other scanning probe techniques to subtract background or enhance resolution \cite{dong2010artifact}, to our knowledge, this is the first demonstration of such resolution enhancement for MIM \cite{lai2009tapping}.

\section{Discussion}

We have presented a new approach to MIM that elevated the state of the art in several aspects. Next, we discuss the promising prospects this work offers, both in enhancing MIM itself and in integrating it with other complementary scanning probe techniques. 

\textit{Spectroscopic mode.} Analogous to the leap from scanning tunneling microscopy to spectroscopy, achieving broadband microwave impedance spectroscopy (MIS) -- which measures the local impedance \textit{spectrum} that allows distinguishing between permittivity and conductivity contributions and investigating narrow-band microwave resonances \cite{wu2017low} -- has been a long-sought goal. However, a major obstacle to continuously tunable broadband operation has been the cancellation circuit, now no longer a necessity. We anticipate that by implementing resistive impedance matching and using length-matched cables for the mixer reference, MIS can be realized with only a moderate sacrifice in SNR \cite{shan2022universal}.

\textit{Nonlinear operation.} Nonlinear MIM (NL-MIM) down-converts the microwave reflection at harmonics of the incident microwave frequency and captures local nonlinear response, critical for understanding many systems of practical and fundamental interest, such as doped semiconductors, superconductors, and correlated materials \cite{ma2016emerging,cho1996scanning,ideue2021symmetry,itahashi2022giant,wang2023microscopic}. The elimination of cancellation, combined with the adoption of broadband active mixers with a wide range of acceptable reference power, significantly simplifies the construction of NL-MIM. To measure across multiple harmonics concurrently, one could simply integrate a series of diplexers and frequency multipliers with a broadband LNA and replicates of the mixer-INA circuit.

\textit{Cryogenic operation.} The cancellation-free architecture is expected to stay Johnson-noise-limited at cryogenic temperatures because a much lower $P_\mathrm{in}$ is typically used to prevent local heating or disturbing delicate samples. For example, at 4 K, the combined Johnson and LNA noise will be $\sim19$ dB lower than that at 295 K, assuming a state-of-the-art cryogenic amplifier. Therefore, for the configuration as in our prototype, the maximum $P_\mathrm{in}$ for Johnson-noise-limited operation will also be reduced by $\sim19$ dB, to -29 dBm. However, this is still higher than the -50 to -30 dBm typically used for cryogenic MIM \cite{lai2011imaging,cui2016unconventional}. Lower source phase noise, higher probe mechanical resonance, or better impedance matching will expand the Johnson-noise-limited regime.

\textit{Higher sensitivity.} Higher sensitivity is attainable by further narrowing the measurement bandwidth. Operating in a controlled environment -- enclosed, temperature-stabilized, or vacuum -- could allow even longer pixel dwell times, albeit with diminishing returns. To achieve even higher sensitivity one likely needs to combine long pixel dwell time with cryogenic temperature, moderate power, a well-matched probe with high responsivity, and a low-phase-noise source. For example, a combination of -150 dBc/Hz source phase noise figure and a responsivity $\eta=$ 7.5 is projected to achieve 0.27 zF sensitivity at 77 K with $P_\text{in}=-10$ dBm and a pixel dwell time of 160 ms (Fig. 4e).  

% M9484C VXG from KeySight

\textit{Multi-modal integration.} The successful utilization of metal-coated monolithic Si probes paves the way to integrate MIM with complementary scanning probe techniques that require (or benefit significantly from) these probes, such as scanning near-field optical microscopy \cite{zhang2013quantitative}, magnetic force microscopy \cite{martin1987magnetic}, and acoustic force microscopy \cite{rabe1994acoustic}. This convergence allows simultaneous multi-modal characterization of complex materials or devices with multiple degrees of freedom. 

\textit{Broader adoption and novel applications.} Our approach significantly lowers the technical and economic barriers to building a high-performance MIM. Detailed designs and resources are made accessible in the Methods section and our repository \cite{Shan2023github}, which we hope will facilitate a broad uptake of MIM in the condensed matter physics and material sciences community. Moreover, since gold-coated silicon probes are widely used to study biological and chemical samples in complex environments \cite{wong1998covalently}, our demonstration extends MIM's applicability to these domains, opening avenues for new insights into a plethora of topics \cite{chu2020microwave,jin2019quantitative,farina2019inverted}.

% , such as near-field optical, magnetic force, and acoustic force microscopy.....Our adoption of Si probes with conductive coating for MIM enables the integration of MIM with other types of scanning probe microscopy. Sensitive force detection methods like magnetic force and acoustic force microscopy require Si-based probes with high $Q$-factors, while near-field scanning optical microscopy benefits from tips fabricated from a conductive Si probe \cite{zhang2013quantitative}. 
% Despite the good MIM performance of our conductive Si probe, it can be further improved by adopting other types of off-the-shelf probes. Silicon nitride probes with one-side reflective gold coating reduces the self-capacitance, and platinum silicide probes offer better wear resistance and less loss. Simultaneous MIM+NSOM/MFM etc... Biological/liquid sample compatibility

% We note that the atomic force microscope (AFM) was operating in the open and none of the components was temperature stabilized. Enclosing the system and adding active temperature stabilization, or operating in a vacuum and/or cryogenic environment will likely make sub-zF sensitivity within reach.  

% Final summary and outlook.... Too long, we probably don't need this 

\section{Methods}

\textit{MIM electronics and probes.} Here we provide the full list of components for our prototype. The microwave source is Signal Hound VSG60A and the LNA is Lotus LNA700M6G2S. The optional microwave spectrum analyzer for monitoring the reflected microwave is Signal Hound SA124B. The directional couplers are Mini-Circuits ZUDC20-0283-S+. The splitter is Mini-Circuits ZFSC-2-10G+. The phase shifter is PNR P1407. We use Moku:Go as LIA and baseband spectrum analyzer. The monolithic Si cantilever probe with an overall gold coating is OPUS 160AC-GG. The SMC probe is Rocky Mountain Nanotechnology 25PtIr200B-H. The SL probe is PrimeNano M300S. 

The design and Gerber files of the impedance matching and mixer-INA PCB can be found in our repository \cite{Shan2023github}. We use a half-wave resonator made of a 100-mm-long RG-178 coax that gives rise to matched frequencies 1 GHz apart. The coax is wire-bonded to the probes on one end and connected to the MIM electronics via a 0.2 pF series capacitor (Johanson S201TL) on the other end. The active mixer is Analog Devices ADL5380 with a 0.5-6 GHz bandwidth. The INAs are Analog Devices AD8428 with a fixed 2000x voltage gain. We limit the INA bandwidth to 900 kHz.

In addition to the benefits described in the main text, our design also eliminates the need for a tunable attenuator to control $P_\text{in}$ and a low-phase-noise amplifier for driving the mixer reference port. The active mixer has a virtually constant conversion gain over a very wide range of reference power, so we can change $P_\text{in}$ simply by changing the source output power through software (Fig. 1a). 

\textit{Scanning parameters.} We used a Park NX7 atomic force microscope (AFM) to perform the scans. We used a state-of-the-art PrimeNano ScanWave Pro as the commercial MIM electronics to benchmark against. We set the time constant of the LIA to be 1/4 of the pixel dwell time, in order to ensure sufficient settling between pixels. Unless otherwise noted, we used a $P_\mathrm{in}$ of -10 dBm and a tapping amplitude of 20 nm. All scans were performed under ambient conditions at 295 K with no enclosure. 

\textit{Feasibility of metal-coated Si probes.} We model the 160AC-GG probe as the probe body in series with the cantilever. In both regions, the resistance from the 70 nm gold coating ($2.44 \times 10^{-8}$ $\Omega\cdot$m, skin depth 1.3 $\mu$m at 3.2 GHz) is in parallel with the highly-doped silicon substrate ($2.5 \times 10^{-4}$ $\Omega\cdot$m, skin depth 0.14 mm). The silicon contribution is small at GHz frequencies due to the skin effect, so can be ignored here. We estimate 0.4 $\Omega$ across the probe body, 0.7 $\Omega$ across the cantilever, and less than 1 $\Omega$ from the tip pyramid, adding up to $\sim2$ $\Omega$, which is confirmed by our measurements. This figure is moderate compared with the 0.2 $\Omega$ of SMC probes and $\sim5$ $\Omega$ of SL probes \cite{yang2012}.

%Because the gold resistance is much smaller and the gold thickness is much smaller than its skin depth, the resistance at 3.2 GHz is expected to remain close to 0.4 $\Omega$.

On the other hand, although Si probes have relatively large unshielded areas, the self-capacitance with a distant ground is estimated and measured (at 100 kHz) to be smaller than 0.2 pF, larger than the $\sim0.03$ pF of SMC or TF probes but less than the $\sim1$ pF of SL probes. 
%At 3 GHz Although the coating thickness of 70 nm is much smaller than the microwave skin depth of 1.4 $\mu$m at $\sim3$ GHz, this is partly mitigated by the smaller resistivity of gold. We measured a probe resistance of $\sim2$ $\mathrm{\Omega}$. Although this is higher than typical resistance values for SMC probecite{shan2022universal}, it has virtually no effect on the microwave voltage delivered to the probe according to our SPICE simulations.

Moreover, with a strategically placed ground plane, much of the capacitance can be compensated by self-inductance to achieve a nearly resistive characteristic impedance close to 50 $\Omega$. To this end, we separated the probe from a ground plane with a 0.5 mm thick fused silica ($\epsilon=3.8$), forming a microstrip waveguide. The characteristic impedance of this microstrip is estimated to be 44 $\mathrm{\Omega}$, close to the system impedance of 50 $\mathrm{\Omega}$. This effectively makes the probe body a short extension of the half-wave resonator coax, instead of a large capacitive load. The single-pass attenuation of this microstrip section is estimated to be 0.1 dB from the resistance values above, smaller than the $\sim$0.25 dB from the RG-178 half-wave resonator coax. 

%\textit{Scaling of signal versus $P_\mathrm{in}$.} The signal strength scales with $P_\mathrm{in}$ (Extended Data Fig. 1), indicating that the microwave response in our system is linear. Therefore, we expect that the SNR scales with $P_\mathrm{in}$ as long as we operate in the thermally-limited regime.

\textit{SNR characterization.} We measure SNR using two complementary approaches: from MIM-Im images and spectrum analyzer measurements. In the first approach, the signal is the contrast between the average MIM-Im values of Al and SiO$_2$ regions in MIM images (Extended Data Fig. 2a), and the noise is the standard deviation of MIM signal within a 10 $\times$ 10-pixel feature-less SiO$_2$ region. In the second method, we use a spectrum analyzer to directly measure the spectra of the MIM-Im channel with the resolution bandwidth (RBW) matching the lock-in time constant. We take the difference in the $f_0$ peak amplitude when the tip is on Al or SiO$_2$ as the signal, and the baseline amplitude slightly away from $f_0$ when the tip is on SiO$_2$ as the noise. We then multiply the SNR extracted from the spectrum analyzer by $2/\sqrt{\pi}$ to account for the conversion factor of the LIA. These two methods yield consistent results (Fig. 3d of the main text).

\textit{Signal versus vibration amplitude.} The dynamic MIM signal also depends on the cantilever vibration amplitude. Extended Data Fig. 2b shows an example, where the contrast between Al and SiO$_2$ first increases then saturates above $\sim17.5$~nm peak-to-peak amplitude, while the raw signals keep increasing. This trend can be understood from the fact that although the dc-MIM versus tip-sample distance curves are monotonic -- hence large vibration amplitude always gives rise to higher dynamic-mode signal, the curves for different materials converge when the tip is far away from the sample, hence the saturation in contrast. We chose 20 nm for an optimal trade-off between MIM signal and topography performance.

\textit{Sensitivity estimation.} To convert the total in-band noise level to a capacitance sensitivity, we used the formula \cite{shan2022universal}$$
    V_{\text{MIM}} = -G\frac{\Delta Y}{2Y_0}\eta^2V_\text{in},
$$ where $G$ is the total gain, $\Delta Y$ is the admittance variation signal, $\eta=$ 7.5 is the measured circuit responsivity, $Y_0=1/(50\Omega)=0.02$ S is the system characteristic admittance, and $V_\mathrm{in}=0.1$ V is the incident microwave voltage for $P_\mathrm{in}=-10$ dBm. To measure $\eta$, we compared our MIM signal contrast between Al and SiO$_2$ to that from a reference system with a 50 $\Omega$ shunt resistor impedance matching which has an $\eta$ of 1. Setting the MIM root-mean-square (rms) signal value $V_{\text{MIM}}$ to be equal to the total in-band rms noise, we can obtain the corresponding $\Delta Y$ and then estimate the capacitive sensitivity $\Delta C=\Delta Y/(2\pi f)$, where $f=3.2$ GHz.

\textit{Encapsulated graphene sample fabrication.} The sample was fabricated using a dry-transfer~\cite{lei_wang2013} method. First, graphene and hBN flakes were exfoliated using tapes onto Si substrates covered by 90-nm thermal oxide. Then, a stamp made of glycol-modified polyethylene terephthalate (PETG) was used to pick up the top hBN, graphene, and bottom hBN in sequence around 70 \degree C.  The stack including the stamp was subsequently released onto a Z-cut quartz substrate at 100 \degree C and soaked in chloroform overnight to dissolve the PETG. Finally, the sample was washed in chloroform, acetone, isopropanol, deionized water, and dried under nitrogen flow. 

\section*{data availability}
The data that support the findings of this study are available from the corresponding author upon reasonable request.

\section*{acknowledgments}
J.-Y.S. and E.Y.M. conceived the project. J.-Y.S., N.M. and E.Y.M. constructed the MIM setup and performed the experiments. S.-D.C. and F.W. fabricated the 2D material sample. J.-Y.S. and E.Y.M. wrote the manuscript with inputs from all authors. This work was supported by the Laboratory Directed Research and Development Program of Lawrence Berkeley National Laboratory under U.S. Department of Energy Contract No. DE-AC02-05CH11231. The graphene sample fabrication is supported by the U.S. Department of Energy, Office of Science, Office of Basic Energy Sciences, Materials Sciences and Engineering Division under contract no. DE-AC02-05-CH11231 (van der Waals heterostructures programme, KCWF16). S.-D.C. acknowledges support from the Kavli ENSI Heising-Simons Junior Fellowship.

\section*{Ethics declarations}
The authors declare no competing interests.

% \section*{References}
\bibliography{cancellation}

% Redefine figure environment and labels for Supplementary Information
\makeatletter
\renewcommand{\fnum@figure}{Extended Data Fig. \thefigure}
\makeatother
\setcounter{figure}{0} % reset counter

\clearpage

\begin{figure}
\includegraphics[width=0.3\textwidth]{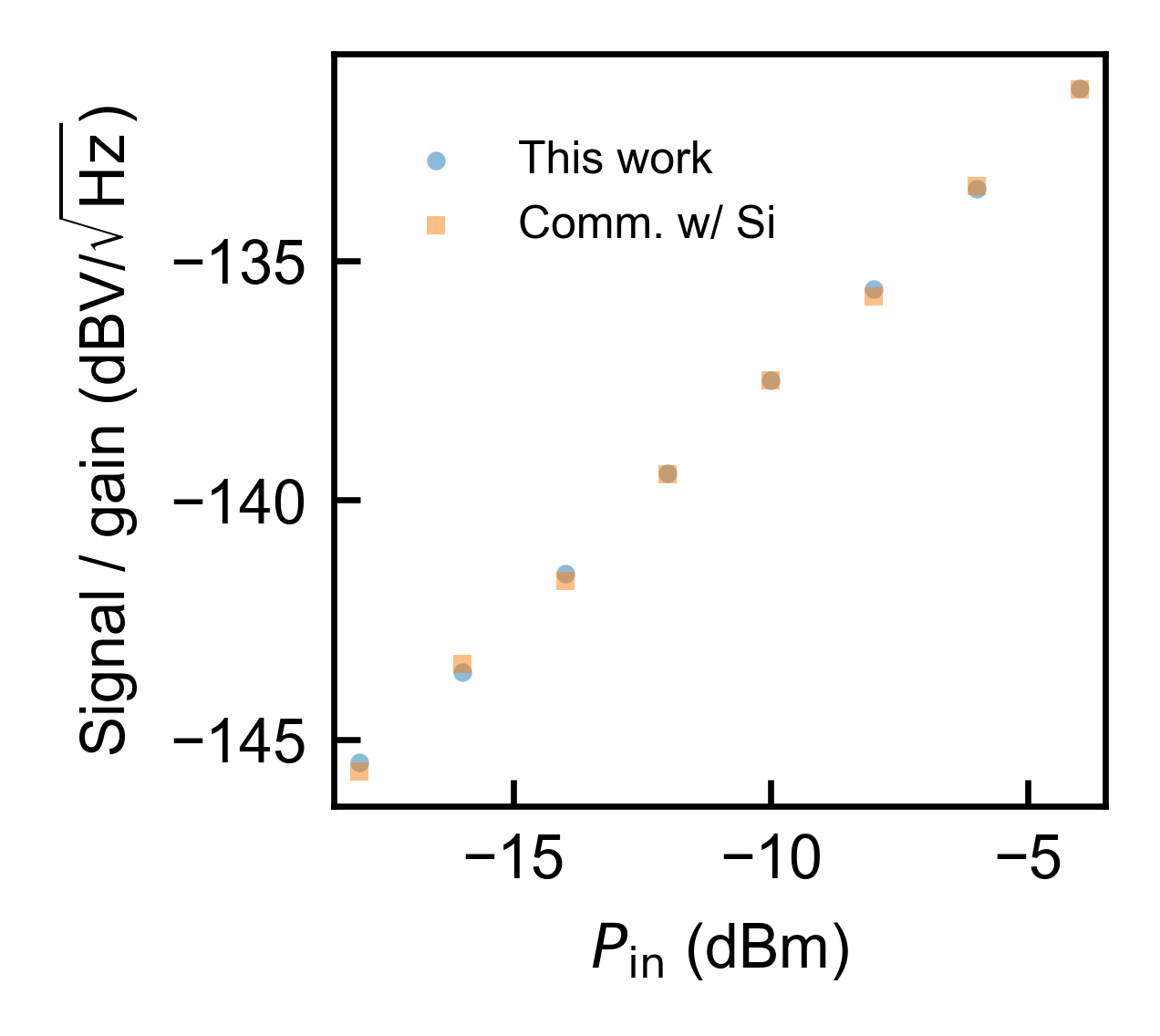}
\caption{\label{extfig2} \textbf{The scaling of dynamic MIM signal with $P_\mathrm{in}$.} Normalized peak amplitude at $f_0$ which represents the dynamic MIM signal, measured as a function of $P_\mathrm{in}$ for our cancellation-free electronics and a commercial system with multi-stage cancellation. The linear dependence shows the absence of saturation in the signal chain.}
\end{figure}

\begin{figure}
\includegraphics[width=0.5\textwidth]{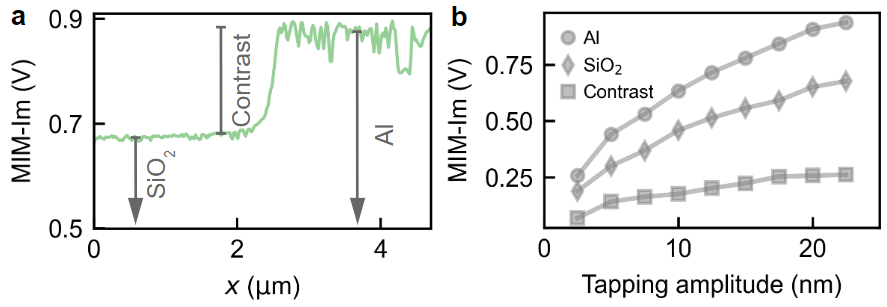}
\caption{\label{extfig1} \textbf{Raw and contrast signal and their dependence on cantilever vibration amplitude. a,} The one-dimensional cut of an MIM-Im image taken on an Al dot sample, showing the raw Al and SiO$_2$ signals as well as their contrast. \textbf{b}, The MIM-Im signal as a function of peak-to-peak cantilever vibration amplitude.}
\end{figure}

\end{document}